\documentclass[aps,pra,reprint,superscriptaddress,10pt,showpacs,a4paper]{revtex4-1}
\usepackage{amsmath,amssymb,amsfonts}
\usepackage{mathtools}
\usepackage{graphicx}
\usepackage{txfonts}
\usepackage{color, xcolor}

\usepackage[unicode,breaklinks]{hyperref}
\hypersetup{
    unicode=true,
    a4paper=true,
    plainpages=false,
    pdftitle={Statics and Dynamics of a Vortex in a Binary Immiscible Bose-Einstein Condensate},
    pdfauthor={R. Doran, A. W. Baggaley, and N. G. Parker},
    pdfsubject={Statics and Dynamics of a Vortex in a Binary Immiscible Bose-Einstein Condensate},
    colorlinks=true,
    linkcolor=blue,
    citecolor=blue,
    filecolor=black,
    urlcolor=blue
}
\usepackage{tikz}

\newcommand{\red}[1]{\textcolor{red}{#1}}

\newcommand{\rr}{\mathbf{r}}

\begin{document}
\title{Vortex Solutions in a Binary Immiscible Bose-Einstein Condensate}

\author{R. Doran\address{ryan.doran@newcastle.ac.uk}}
\affiliation{Joint Quantum Centre (JQC) Durham--Newcastle, Department of Mathematics, Statistics and Physics, Newcastle University, Newcastle upon Tyne, NE1 7RU, UK}

\author{A. W. Baggaley}
\affiliation{Joint Quantum Centre (JQC) Durham--Newcastle, Department of Mathematics, Statistics and Physics, Newcastle University, Newcastle upon Tyne, NE1 7RU, UK}

\author{N. G. Parker}
\affiliation{Joint Quantum Centre (JQC) Durham--Newcastle, Department of Mathematics, Statistics and Physics, Newcastle University, Newcastle upon Tyne, NE1 7RU, UK}

\date{\today}

\begin{abstract}
We consider the mean-field vortex solutions and their stability within a two-component Bose Einstein condensate in the immiscible limit. A variational approach is employed to study a system consisting of a majority component which contains a single quantised vortex and a minority component which fills the vortex core. We show that a super-Gaussian function is a good approximation to the two-component vortex solution for a range of atom numbers of the in-filling component, by comparing the variational solutions to the full numerical solutions of the coupled Gross-Pitaevskii equations. We subsequently examine the stability of the vortex solutions by perturbing the in-filling component away from the centre of the vortex core, thereby demonstrating their stability to small perturbations.
\end{abstract}

\maketitle

\section{Introduction}
\label{section:introduction}
Quantized vortices have been realised in a variety of superfluid systems, including superfluid $^{4}\mathrm{He}$, phases of superfluid $^{3}\mathrm{He}$, and atomic Bose Einstein condensates (BECs) \cite{Barenghi_review_2014}; the latter was first achieved by using a phase imprinting method on a two-component BEC \cite{Matthews1999}. Since this initial observation in BECs, vortices have been realised in a number of scenarios including rotating single \cite{Chevy2000,Raman2001} and multiple \cite{Schweikhard2004} component BECs, nucleation from a repulsive potential \cite{Inouye2001,Kwon2015,Kwon2016}, and driving the condensate out of equilibrium in harmonic \cite{Henn2009_PRA,Henn2009_PRL} and uniform \cite{Navon2016,Navon2019} potentials. Of particular interest are the latter experiments\cite{Henn2009_PRA,Henn2009_PRL,Navon2016,Navon2019}, where the turbulent system manifests itself as many vortices distributed in a complex and disordered tangle.

In superfluid helium, where the typical size of a vortex core is of the order of $0.1\mathrm{nm}$ \cite{Bewley2006}, a range of methods have been theoretically and experimentally evaluated as ways to visualise the flow and track vortices. Many of these techniques involve the use of tracer particles, micron-sized solid particles which are suspended in the fluid and follow the local fluid velocity \cite{Chopre1957,Chung1965,Zhang2005,Bewley2006,Bewley2008,Guo2009}. A significant drawback of these techniques is that the particles are much larger than the vortices which they track, prohibiting detailed probing of quantum turbulence \cite{Guo2014}. By comparison, the most common way to image quantized vortices in a BEC is to allow the condensate to expand until the size of the vortex cores exceed the optical resolution limit and perform column-integrated imaging of the cloud \cite{Madison2000,Raman2001}. Advances in this technique now allow for imaging in multiple dimensions \cite{Donadello2014}, and in real time \cite{Freilich2010,Serafini2015}. However, it remains an ongoing challenge to image a complex 3D distribution of many tangled vortex lines.   

Binary BECs, consisting of two co-existing condensate components, have been achieved experimentally with two hyperfine states of the same atomic species \cite{Myatt1997,Hall1998,Matthews1999,Maddaloni2000,Delannoy2001,Schweikhard2004,Mertes2007,Anderson2009,Tojo2010,Miesner1999}, different isotopes of the same atomic species \cite{Papp2008}, and with different atomic species \cite{Ferrari2002,Modugno2002,Thalhammer2008,McCarron2011}. The two components are coupled, and the components may be miscible or immiscible, depending on the inter-species and intra-species interaction strengths \cite{Pu1998}. Compared to single-component BECs, the coupling gives rise to an exotic array of steady state solutions \cite{Pu1998,Ho1996,Timmermans1998,Ao1998,Trippenbach2000,Barankov2002,VanSchaeybroeck2008,Guatam2010,Gordon1998,Kim2002}, from overlapping density profiles to phase-separated profiles. Since their initial realisation \cite{Matthews1999}, vortices have been studied in two-component condensates both experimentally \cite{Schweikhard2004,Anderson2000} and theoretically \cite{Feder1999,Garcia-Ripoll2000,Chui2001,Jezek2001,Ohberg2002,Park2004,Woo2007,Kasamatsu2009,Yakimenko2009,Catelani2010,Law2010,Kuopanportti2012}. More recently, theoretical work has concentrated on the relaxation of a turbulent two-component BEC \cite{Mithun2021,Wheeler2021}, dynamics of vortices in a two-component BEC \cite{Li2019,Han2019,Richaud2020,Richaud2021}, and the importance of the cross-over between the miscible and immiscible regimes \cite{Bandyopadhyay2017,He2019}.   

In this paper, we concentrate on a  two-component BEC which is in the immiscible regime and where one component contains the majority of the total atoms.  A vortex in the majority component is then known to become in-filled by the second component \cite{Matthews1999,Anderson2000}. Such a regime is a potential candidate for three-dimensional vortex detection in systems comprising complex vortex tangles, as the in-filling component can be tracked without destructively imaging the majority component. It is possible, however, that the presence of the in-filling component will modify the vortex states and their dynamics, and that this modification will primarily depend on the number of in-filling atoms and the inter-species interactions. It is important for the purpose of vortex detection to understand the regimes in which the in-filling component has no significant effect on the vortex dynamics, that is, is an effective passive tracer.  It is also interesting to consider how, in more extreme cases, the in-filling might drive new regimes of vortex states and behaviour which have no analogy in conventional single-component superfluids. In this paper we thus investigate the in-filled vortex solutions in an immiscible binary BEC, establishing their profiles through a variational approach and full numerical approaches, as well as their stability under perturbation.

The remainder of this paper is structured as follows: in Section \ref{section:equation_of_motion} we introduce the equations of motion of the system, the coupled Gross-Pitaevskii Equation. In Section \ref{section:Statics} we derive a variational approach to the in-filled vortex solutions, which allow the solutions to be established analytically. The variational solutions are compared to numerically-obtained solutions of the coupled Gross-Pitaevskii equation, showing excellent agreement. Section \ref{section:dynamics_single} establishes the stability of the in-filled vortices to small perturbations, before we present concluding remarks in Section \ref{section:conclusions}.


\section{The Coupled Gross-Pitaevskii Equation}
\label{section:equation_of_motion}

A system which consists of two BECs which are dilute and weakly-interacting in the zero-temperature limit is well described by the coupled Gross-Pitaevskii Equation (CGPE) \begin{subequations} 
\begin{align}
    i\hbar \frac{\partial \psi_1}{\partial t} &=& \left[ - \frac{\hbar^2}{2m_1} \nabla^2 + V_1 \left(\rr\right) + g_{11} |\psi_1|^2 + g_{12} |\psi_2|^2\right] \psi_1, \label{eqn:cgpe1} \\
    i\hbar \frac{\partial \psi_2}{\partial t} &=& \left[ - \frac{\hbar^2}{2m_2} \nabla^2 + V_2 \left(\rr\right) + g_{12} |\psi_1|^2 + g_{22} |\psi_2|^2\right] \psi_2,
    \label{eqn:cgpe2}
\end{align}
\end{subequations}
where $\psi_k=\psi_k(\rr,t)$ is the mean-field wavefunction of the $k$th component, $k=1,2$, and $m_k$ is the mass of the atomic species in the $k$-th component. Each of the components are independently subject to an external trapping potential $V_k(\rr)$. The intra-species interactions of the first and second components are parameterised by $g_{11}$ and $g_{22}$ respectively, while the inter-species interactions are given by $g_{12}$. The density profile of each component is $n_1(\rr,t)=|\psi_1(\rr,t)|^2$ and $n_2(\rr,t)=|\psi_2(\rr,t)|^2$, and each component is normalised to $N_1$ and $N_2$ atoms.

In what follows, we will consider species 1 to be the majority component, and refer to species 2 as the in-filling component. For simplicity, we set the external potentials to zero, such that the ground-state of the majority component is a state of uniform density.  Then, it is useful to work in natural units of the majority component: density is in terms of the uniform density $n_{0,1}$, length is in terms of the healing length, $\xi_1 = \hbar/\sqrt{n_{0,1} m_1 g_{11}}$, energy is given by the chemical potential, $\mu_1 = n_{0,1} g_{11}$, and time is given by $\tau_1 = \hbar/\mu_1$. The CGPE, Eqns.~\eqref{eqn:cgpe1} and \eqref{eqn:cgpe2}, may then be cast in dimensionless form
\begin{subequations}
\begin{align}
    i \frac{\partial \psi_1'}{\partial t'} &=& \left[ \quad - \frac{1}{2} \nabla^{\prime 2}  + \ \quad |\psi_1'|^2 + g_{12}' |\psi_2'|^2 \right] \psi_1', \label{eqn:dimensionless_cgpe1} \\
    i \frac{\partial \psi_2'}{\partial t'} &=& \left[ - \frac{1}{2} m' \nabla^{\prime 2} + g_{12}' |\psi_1'|^2 + g_{22}' |\psi_2'|^2 \right]\psi_2', \label{eqn:dimensionless_cgpe2}
\end{align}
\end{subequations}
where the dimensionless parameters are the ratio of the atomic masses, $m'=m_1/m_2$, the ratio of the inter- and intra-species interaction parameters, $g_{12}' = g_{12}/g_{11}$, and the ratio of the inter-species interaction parameters, $g_{22}' = g_{22}/g_{11}$. In the following, we will consider a two-component system which is in the immiscible limit,
\begin{equation}
    g_{12}^2 > g_{11} g_{22}, 
\end{equation}
corresponding to $\left(g_{12}'\right)^2>g_{22}'$ in dimensionless variables. For convenience, we will drop the primes in the remainder of this paper.

\section{Numerical and variational approaches to the vortex solution}
\label{section:Statics}
\subsection{Overview}
We consider an in-filled vortex which is aligned along the $z$-axis of the system, and aim to determine its cross-sectional profile.  Since the vortex is assumed to be uniform along $z$, this reduces to a two-dimensional problem.  The core of a vortex is characterised by a region of depleted density containing a point of zero density, around which the phase winds by an integer multiple of $2\pi$ \cite{Pethick_and_Smith}. In the immiscible regime, it is energetically favourable for the minority component to be located at regions where the density of the majority component is lower\cite{Trippenbach2000}. Therefore, if the majority component contains a vortex, we expect that the minority component ``infills'' the vortex core. 

Throughout the remainder of the paper we will set $g_{22}=1.0$ and $g_{12}=1.1$ unless otherwise stated, meaning that the two species are just in the immiscible regime. We also take the ratio of masses to be equal, corresponding to two components which are made up of identical atomic species in different hyperfine states.

First we outline our approach to obtaining the full numerical solutions before establishing the semi-analytic variational approach.

\subsection{Numerical solutions}
We obtain the full vortex solutions by numerically solving the CGPE; these solutions will allow us to later test the success of the variational approach.

We solve the CGPE in two-dimensions using an adaptive RK45 method, with an error tolerance of $10^{-10}$,  implemented using XMDS2 \cite{XMDS2}.  We do so in a square computational grid with 2 numerical grid points per healing length, up to $1000$ time-steps.  In order to obtain the vortex solutions we impose an azimuthal phase on to the majority component, and evolve the CGPE, Eqns.~\eqref{eqn:dimensionless_cgpe1} and \eqref{eqn:dimensionless_cgpe2}, under a Wick rotation $t\to i \tau$.  Under this so-called imagingary time propagation \cite{Barenghi2016}, the family of GPEs are well established to evolve towards the lowest energy state of the system.  To avoid issues with the phase at the edge of the (periodic) domain, we impose a circular hard-wall potential with radius $R$ and amplitude $100\mu$ in each component to force its density to zero; we have checked that the radius $R$ is sufficiently large that this imposed boundary has no effect on the in-filled vortex solutions.

\begin{figure}
    \centering
    \includegraphics{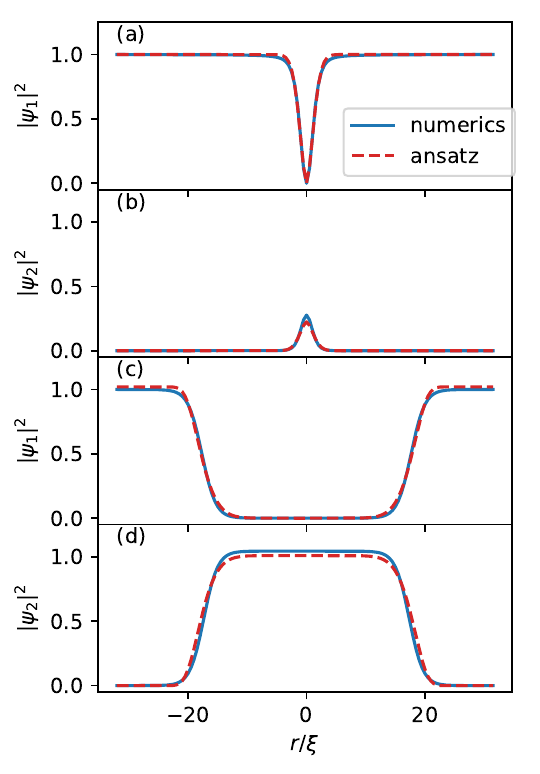}
    \caption{Example density profiles of the in-filled vortex solutions for (a),(b) $N_2=2$ and (c),(d) $N_2=1000$. In panels (a) and (c) the blue solid curves show the numerically obtained density profile of the majority component, $|\psi_1|^2$,  while the red dashed curves show the density profile given by the variational solution. In panels (b) and (d) the blue solid curve shows the numerically obtained density profile of the minority component, $|\psi_2|^2$, while the red dashed curve show the corresponding variational solution. In each case, $g_{12}=1.1$, $g_{22}=1.0$ and $m'=1.0$.}
    \label{fig:example_solutions}
\end{figure}

Example numerical solutions from the CGPE are shown in Fig.~\ref{fig:example_solutions}.  For a small minority component (panels (a, b)), the vortex density profile in the majority component resembles that of a vortex in a single-component BEC \cite{Barenghi2016}; the density is zero at the centre of the vortex and relaxes to the background density value over a lengthscale characterised by the healing length $\xi$.  The minority component is localised within the vortex core as a narrow wavepacket whose width is consistent with the vortex core, i.e. the healing length $\xi$.   For a large minority component (panels (c,d)), however, the vortex profile in the majority component is much broader and flat-bottomed, with the minority component forming a broad flat-topped profile of similar width.

Physically, it is evident that, due to the immisiciblity of the two components, it is energetically favourable for the minority component to sit at the vortex core so as to minimise overlap of the two components.  As more atoms are added to the minority component, these cause the vortex core to broaden, again so as to avoid overlap of the two components.

\subsection{Variational solutions using a super-Gaussian ansatz}

It is our aim here to establish a semi-analytic approach to the in-filled vortex solutions, for both components, using a variational method.  
Variational methods have been employed to find stationary solutions of a variety of systems, including vortex cores in a single-component BEC \cite{Pethick_and_Smith,Bradley_Anderson2012}, bright solitons in BECs of attractive \cite{Perez1998,Carr2002,Salasnich2002,Parker_2007,parker_2009,Billam2012} and dipolar \cite{Edmonds2017} atomic species, quantum droplets in vanilla \cite{Otajonov2020,Lavoine2021} and dipolar BECs \cite{Poli2021}, and bosonic quantum impurities \cite{Edmonds2021}. The main advantage of these variational methods is the relative ease with which stationary solutions may be obtained, by comparison with the computational requirements of finding solutions to the full CGPE, and the results can often give useful physical insight into the properties of the solutions \cite{Barenghi2016}.

It can be shown that the energy functional corresponding to Eqns.~\eqref{eqn:dimensionless_cgpe1} and \eqref{eqn:dimensionless_cgpe2} may be written as
\begin{eqnarray}
    E\left[\psi_1,\psi_2\right] &=& \int \ d^2\rr \ \left[ \frac{1}{2} |\nabla\psi_1|^2 + \frac{1}{2} m |\nabla\psi_2|^2\right] \nonumber \\
    &+& \int \ d^2 \rr \left[ \frac{1}{2} |\psi_1|^4 + g_{12}|\psi_1|^2 |\psi_2|^2 + \frac{1}{2} g_{22}|\psi_2|^4 \right], \nonumber \\
    \label{eqn:energy_functional}
\end{eqnarray}
where 
\begin{equation}
    i \frac{\partial \psi_k}{\partial t} = \frac{\delta E}{\delta \psi_k^*}
\end{equation}
for $k=1,2$. 

In this section we proceed by substituting an ansatz solution into the energy functional, Eqn.~\eqref{eqn:energy_functional}; this results in an energy function which depends on parameters controlling the density profile of the two components. We minimise this function to find the variational solution. 

The choice of ansatz for each component must satisfy the following limits. Firstly $|\psi_1(r)|^2 \to n_{0,1}$ and $|\psi_2(r)|^2 \to 0$ as $r\to\infty$, which is to say that, well away from the vortex core, the density of the majority component relaxes to the density of the uniform background and the density of the minority component reduces to zero. Secondly, $|\psi_1(r)|^2 \to 0$ and $|\psi_2(r)|^2 \to \psi_{2,max}^2$ as $r\to 0$, where $\psi_{2,max}^2$ is the peak density of the minority component.  Moreover, the ansatz should be able to capture the range of profiles illustrated in Fig. \ref{fig:example_solutions} - from narrow, high-curvature profiles to broad, flattened profiles. Here we choose to base our ansatz on the super-Gaussian functions, which is sufficiently versatile to satisfy these criteria.

We write the ansatz solution for the in-filled vortex in polar coordinates $(r,\varphi)$ as 
\begin{equation}
    \psi_1 = A \left\{ 1 - \exp\left[-\left(\frac{r}{\lambda}\right)^{2\alpha}\right] \right\}^{1/2} \exp\left(iq\varphi\right)
    \label{eqn:majority_ansatz}
\end{equation}
and
\begin{equation}
    \psi_2 = B \exp\left[-\frac{1}{2}\left(\frac{r}{\lambda}\right)^{2\alpha}\right],
    \label{eqn:infill_ansatz}
\end{equation}
where $q$ is the charge of the vortex, $\lambda$ is a parameter which characterizes the width of the vortex core and in-filling component.  The exponential terms are the super-Gaussian function, and therein the exponent $\alpha$ controls the shape of the function. The pre-factors
\begin{equation}
    A = N_1^{1/2} \left\{ \pi R^2 - \frac{\pi \lambda^2}{\alpha} \Gamma\left[\frac{1}{\alpha},\left(\frac{R}{\lambda}\right)^{2\alpha}\right]\right\}^{-1/2}, 
\end{equation}
and 
\begin{equation}
    B = \left(\alpha N_2\right)^{1/2} \left\{ \pi \lambda^2 \Gamma \left[\frac{1}{\alpha}, \left(\frac{R}{\lambda}\right)^{2\alpha}\right]\right\}^{-1/2}
\end{equation}
normalise the ansatz wavefunctions to $N_1$ and $N_2$ respectively,
where $\Gamma\left(s,x\right)$ is the incomplete Gamma function \cite{Abramowitz1965handbook}, defined in Eqn.~\eqref{eqn:incomplete_gamma_function}. According to the super-Gaussian function, for $\alpha<1$ we obtain a cusp profile, while
for $\alpha=1$ we recover a vanilla Gaussian curve,  and for $\alpha>1$ the curve is a flat-topped. Example fits of this ansatz to numerical solutions of the CGPE can be found in Fig.~\ref{fig:example_solutions}; clearly we see that the ansatz suitably captures both the narrow, high-curvature and the broad, flattened profiles presented.


We proceed analytically by substituting Eqns.~\eqref{eqn:majority_ansatz} and \eqref{eqn:infill_ansatz} into the energy functional, Eqn.~\eqref{eqn:energy_functional}, and integrating out the spatial dependence. This is a non-trivial calculation and further details can be found in the Appendix.  To prevent the integrals involving the majority component from diverging, we consider the energy of the atoms within a finite distance $R$ of the vortex core, and without loss of generality we place the vortex core at the origin \cite{Pethick_and_Smith}. The resulting equation for the energy functional is 
\begin{eqnarray}
E\left(\lambda,\alpha\right) &=& \frac{\pi A^2 \alpha}{2} \left\{ \left(1+S\right)\exp\left(-S\right) + \mathrm{Sp}\left[\exp\left(-S\right)\right] - 1 \right\} \nonumber \\ 
&+& \frac{\pi q^2 A^2}{2\alpha}\left[ \log\left(S\right) - \mathrm{Ei}\left(-S\right) + \gamma_\mathrm{EM} \right] \nonumber \\
&+&  \frac{\pi m \alpha^2 B^2}{2} \left[ 1- \left(1+S\right) \exp\left(-S\right)\right] \nonumber \\
&+& \frac{\pi A^4}{2}\left[ R^2 - \frac{2\lambda^2}{\alpha} \Gamma\left(\frac{1}{\alpha},S\right) + \frac{\lambda^2}{2^{1/\alpha}\alpha} \Gamma\left(\frac{1}{\alpha},2S\right)\right] \nonumber \\
&+& \frac{\pi g_{12} \lambda^2 A^2 B^2}{\alpha} \left[  \Gamma\left(\frac{1}{\alpha},S\right) - \frac{1}{2^{1/\alpha}} \Gamma\left(\frac{1}{\alpha}, 2S\right)\right] \nonumber \\
&+& \frac{\pi g_{22} B^4 \lambda^2}{2^{1 + 1/\alpha} \alpha}\Gamma\left(\frac{1}{\alpha},2S\right),
\label{eqn:ansatz_energy_functional}
\end{eqnarray}
where $S=(R/\lambda)^{2\alpha}$, and we have introduced the Spence function, $\mathrm{Sp}$, defined in Eqn.~\eqref{eqn:Spence_integral},  the Exponential integral, $\mathrm{Ei}$, defined in Eqn.~\eqref{eqn:Exponential_integral}, and the Euler-Mascheroni constant, $\gamma_\mathrm{EM}\approx0.5772$.

While approximations may exist, such that we can minimise Eqn.~\eqref{eqn:ansatz_energy_functional} analytically, we choose to numerically minimise the energy function as it appears in  Eqn.~\eqref{eqn:ansatz_energy_functional}. In order to do this, we apply the quasi-Newton method of Broyden, Fletcher, Goldfarb and Shanno \cite{BFGS_minimisation}, available in the \textit{scipy} Python library. Thus, for given atom numbers $N_1$ and $N_2$, the variational solution for the in-filled vortex is specified by the two parameters $(\alpha^*,\lambda^*)$ which minimises the variational energy.  These solutions agree well with the full numerical solutions, as evident in Fig. \ref{fig:example_solutions}.

\begin{figure}
     \centering
     \includegraphics{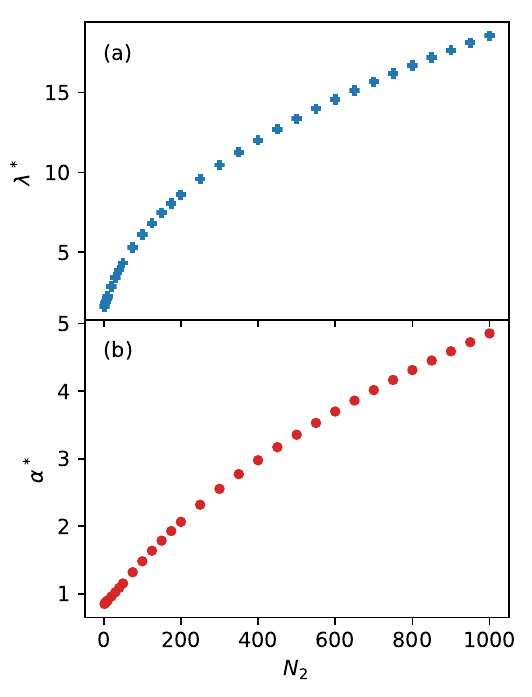}
     \caption{Parameters which minimise the energy of the ansatz, Eqn.~\eqref{eqn:ansatz_energy_functional}, as the number of atoms in the in-filling component, $N_2$, varies. Panel (a) contains the energy-minimizing width, $\lambda^*$, while panel (b) contains the energy-minimizing exponent of the super-Gaussian function, $\alpha^*$. These solutions have scaled intra-species interaction $g_{22}=1.0$, inter-species interaction $g_{12}=1.1$, and mass ratio $m=1.0$.}
     \label{fig:ansatz_parameters}
 \end{figure}

The variation of the variational solution parameters $(\alpha^*,\lambda^*)$ with the number of atoms in the in-filling component, $N_2$ is shown in Fig.~\ref{fig:ansatz_parameters}.  For low $N_2$, the values of $\lambda^*$ and $\alpha$ are small, giving rise to narrow, high-curvature profiles such as in Fig. \ref{fig:example_solutions}(a,b). As $N_2$ is increased, the values of $\lambda^*$ and $\alpha^*$ grow, indicating the broadening and flattening of the profiles in both components, such as the profiles in Fig. \ref{fig:example_solutions}(c.d). 

\subsection{Accuracy of the variational solution}

We now perform a quantitative assessment of the accuracy of the varational solution for the in-filled vortex compared to the full numerical solution of the CGPE. First we consider the energy of the solution. We compute the normalised error in the energy, 
\begin{equation}
    \Delta E = \frac{E\left(\lambda^*,\alpha^*\right) - E_\mathrm{CGPE}}{E_\mathrm{CGPE}},
    \label{eqn:normalised_energy_error}
\end{equation}
where $E(\lambda^*,\alpha^*)$ is the energy of the variational solution (the energy functional Eqn.~\eqref{eqn:ansatz_energy_functional} evaluated at $\left(\lambda^*,\alpha^*\right)$) and $E_\mathrm{CGPE}$ is the energy of the numerical solution. We also compute the normalised maximum deviation in the density of the variational solution from the numerical solution, 
\begin{equation}
    \Delta |\psi_k|^2 = \frac{\max \left| |\psi_{k,\mathrm{Ansatz}}|^2 - |\psi_{k,\mathrm{CGPE}}|^2 \right|}{\max |\psi_{k,\mathrm{CGPE}}|^2 },
    \label{eqn:normalised_density_error}
\end{equation}
where $\psi_{k,\mathrm{Ansatz}}$ is the $k$-th component of the variational wavefunction, and $\psi_{k,\mathrm{CGPE}}$ is the $k$-the component of the numerical wavefunction. In order that this statistic is not affected by the implementation of the circular hard-wall potential in the numerical solution, we compute this in the region $r<R/2$, well away from the hard walls. To judge the goodness of the shape of the variational wavefunction, we also calculate the normalised error in the Full Width at Half Maximum (FWHM) of the density profile, which is given by
\begin{equation}
    \Delta \mathrm{FWHM} \left( |\psi_k|^2\right) = \frac{\left| \mathrm{FWHM}\left(|\psi_{k,\mathrm{Ansatz}}|^2\right) - \mathrm{FWHM}\left(|\psi_{k,\mathrm{CGPE}}|^2\right) \right| }{\mathrm{FWHM}\left(|\psi_{k,\mathrm{CGPE}}|^2\right)}.
    \label{eqn:normalised_FWHM_error}
\end{equation}

\begin{figure}
     \centering
     \includegraphics{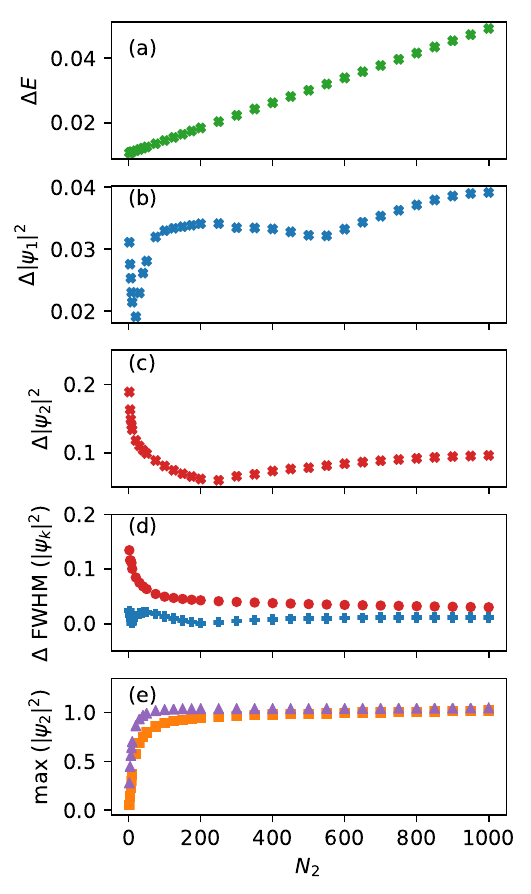}
     \caption{Comparison of the numerically obtained ground state, and the ansatz wavefunction predicted by minimising the energy function, Eqn.~\eqref{eqn:ansatz_energy_functional}. Panel (a), we plot the normalised error between the energy of the ansatz wavefunctions and the energy of the numerically obtained ground state, Eqn.~\eqref{eqn:normalised_energy_error}. We plot the normalised maximum deviation of the ansatz wavefunction from the ground state wavefunction, Eqn.~\eqref{eqn:normalised_density_error}, for the density of the majority [panel (b)] and the in-fill [panel (c)] components. In panel (d), we plot the normalised error in the FWHM, Eqn.~\eqref{eqn:normalised_FWHM_error}, for the majority, $k=1$, component (blue pluses), and the in-fill, $k=2$, component (red circles). In panel (e), we plot the maximum values of the in-filling density for both the ansatz (orange squares) and the numerical groundstate (purple triangles).}
     \label{fig:goodness_of_ansatz}
 \end{figure}

The results of these metrics is plotted in Fig.~\ref{fig:goodness_of_ansatz}. We see that the energy of the variational solution is accurate to within $5\%$ of the numerical solutions throughout the full range of $N_2$ considered.  In panel (c) we note that the normalised maximum error in the density of the in-filling component is largest for small $N_2$. We suggest that this is due to the fact that the maximum value of the in-filling component is relatively small here [see panel (e)], which causes the normalised error to grow quickly. We observe that, while the variational solution for the in-filling component under-estimates the peak density, this error is mainly symptomatic of small $N_2$, and for larger $N_2$ the maximum value of the in-filling density is in good agreement. 

Of particular note is the close agreement between the variational solution and numerical solution for the majority component. This can be seen both in the deviation of the density profiles [panel (b)], and in the deviation of the FWHM [panel (d), blue pluses]. The main motivation of this work is to establish how a second component might affect the ground state of a majority component which contains a vortex.  With this in mind, we might regard the excellent agreement of the variational solution and the numerical solution in the majority component as being more important than the good agreement of the in-filling component.

\section{Response of the vortex to perturbation}
\label{section:dynamics_single}
\subsection{Overview}
Until now we have concentrated on calculating the stationary state of the in-filled vortex. We now turn to considering how stable the vortex solution is to perturbation.  Specifically we will consider the response to perturbing the in-filling component and how localised the minority component remains within the vortex core; this is particularly relevant when considering the possibility to use the minority component as a tracer of vortices.  

\subsection{Perturbing the in-filling component}

We prepare the perturbed state by forming the in-filled vortex solution as previously but then instantaneous translate the in-filling component by a distance $x_0$ along the $x$-axis, relative to the vortex core.  We then evolve this system using Eqns.~\eqref{eqn:cgpe1} and \eqref{eqn:cgpe2} in real time.  We consider in-filling components with two different atom numbers, $N_2=10$ and $N_2=100$. The number of atoms in the majority component, $N_1$ is again chosen so that, far away from the trapping potential or the vortex core, the density is unitary. For the systems which we consider, the number of atoms in the majority component is approximately $5\times 10^3$ and $5\times 10^2$ larger than the number of atoms in the in-filling component, for $N_2=10$ and $N_2=100$ respectively. 


\begin{figure}
    \centering
    \includegraphics{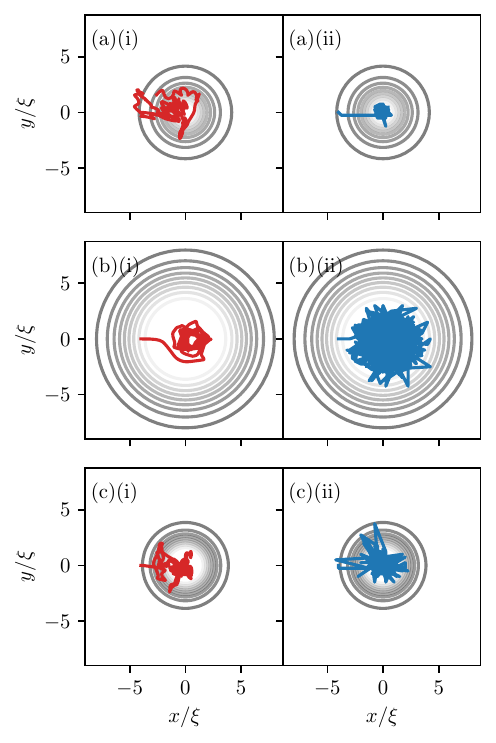}
    \caption{Example trajectories of the in-filling component which is perturbed from the centre of a vortex core in the majority component. Column (i) shows the trajectory of the centre of mass of the infilling component, while column (ii) shows the trajectory of the peak of the in-filling component's density. In row (a), we consider a system where $g_{12}=1.1$ and $N_2=10$; in row (b) we consider a system where $g_{12}=1.1$ and $N_2=100$; in row (c) we consider a system where $g_{12}=2.2$ and $N_2=10$. In all cases, $g_{22}=1.0$ and $m=1.0$.
    The location and size of the vortex in the majority component is indicated by the contour lines; the lines are separated by one-tenth of the background density of the majority component. In each case, the centre of the in-filling component is perturbed by $4\xi$.}
    \label{fig:example_trajectories}
\end{figure}

We track the location of the in-filling component through two approaches.  Firstly, we consider the centre of mass of the component; we define this as
\begin{equation}
    \rr_2 = \frac{1}{N_2} \int |\psi_2|^2 \rr \ d^2 \rr.
\end{equation}
Secondly, 
we track the peak of the in-filling condensate's density, $\rr_2 = \max\left(|\psi_2|^2\right)$. Example trajectories are plotted in Fig.~\ref{fig:example_trajectories} for an initial perturbation of $4\xi$, and for three different parameter sets - (a) $N_2=10$ and $g_{12}=1.1$, (b) $N_2=100$ and $g_{12}=1.1$, and (c) $N_2=10$ and $g_{12}=2.2$. In all three of the cases considered, we observe that the in-filling component is stable to small perturbations away from the vortex core. One would expect this since, as the two components are immiscible, it is energetically favourable for the peak density of the minority component to be attracted to the minimum density in the majority component, i.e. the vortex core. We see that this is true over a wide range of in-filling atom numbers, as well as a range of inter-species interaction strengths. In all cases the in-filling component undergoes an irregular trajectory in the $x-y$ plane; this is due to the non-trivial potential it experiences from the majority component. 

We see that, in the weakly immiscible system ($g_{12}=1.1$) shown in rows (a) and (b), the centre of mass of the system with a larger number of in-filling atoms is subjected to a more tightly confined trajectory. This is because the overlap interaction term ($g_{12}|\psi_1|^2|\psi_2|^2$) in the energy functional, Eqn.~\eqref{eqn:energy_functional}, grows faster with larger $N_2$. Similarly, we find that for systems with a comparable number of atoms [$N_2=10$ in rows (a) and (c)], the centre of mass of the system with the stronger inter-species interaction strength undergoes a more constrained orbit. In rows (b) and (c) we observe that the trajectory of the peak density fluctuates more from the vortex core than the trajectory of the centre of mass. We suggest that, by comparison with row (a), the effective trap which the in-filling component experience (from the interaction potential with the majority component) has a larger radius and shallower gradient in the centre; thus there is more ``sloshing'' of the in-filling component, leading to greater variance in the position of the density peak.  We observed that the vortex undergoes a negligible translation from the origin after perturbing the in-filling component. The fact that this translation is very small is due to the large imbalance between the number of atoms in each component.  

\begin{figure}
    \centering
    \includegraphics{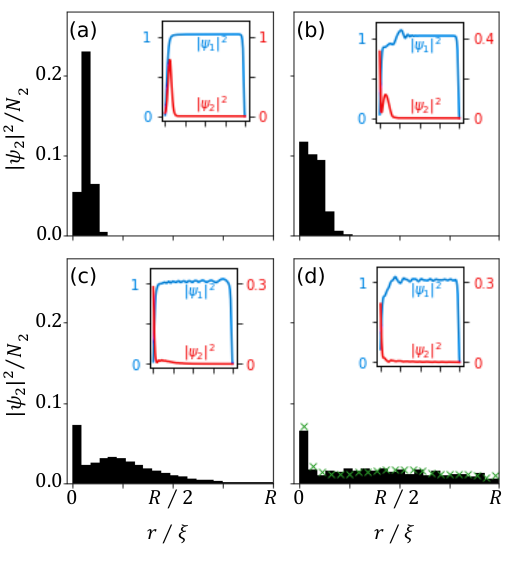}
    \caption{Normalized histogram showing the radial distribution of the infilling component density, $|\psi_2|^2$, after an initial perturbation of $4\xi$ from the vortex core, for $R=64\xi$. Panels represent different snapshots in time: (a) $t=0$, (b) $t=10$, (c) $t=60$, and (d) $t=120$. 
    Green crosses in panel (d) show the time-averaged distribution, averaged over the latter half of the simulation, $500\leq t \leq 1000$. Insets show the density profile of $\psi_1$ (blue), and $\psi_2$ (red), at corresponding times.}
    \label{fig:infill_histogram}
\end{figure}

We trace the coarse-grained density of the in-filling component in Fig.~\ref{fig:infill_histogram}. The effect of instantaneously perturbing the in-filling component away from the vortex core is to generate sound waves, which propagate from the edge of the vortex core. Since the wave-front is an area where the density of the majority component is depleted (although it is non-zero, unlike the vortex core) it carries a small amount of the in-filling component away from the centre of the vortex core. This wave-front collides with the hard-wall trapping potential, and is reflected back into the centre of the trap, interfering with itself. A result of this is that the density waves away from the vortex core have velocities which are radially both inward and outward. Over time, this leads to a redistribution of the in-filling component: while a large amount of the in-filling component remains within the vortex core throughout the simulation, the small amount which is displaced approaches a radial distribution which is approximately uniform [see Fig.~\ref{fig:infill_histogram} (g)-(i)].
It is clear that,  at late times, the majority of the in-filling component remains strongly localised within the vortex under perturbation, and that it's possible suitably traces the position of the vortex core. \footnote{An example movie of this evolution is available in the supplementary material. \red{Link to be added by publisher.}}

Despite the fact that the in-filling component is only perturbed in the $x$ direction, we also observe motion in the $y$ direction. This is due to the fact that, in the majority component, the vortex imposes a velocity field about the origin. We may consider this superfluid velocity field as the velocity at which a particle would be advected \cite{Pethick_and_Smith}, and hence, due to the small overlap between the two components, the in-filling component is subjected to a velocity field in the $x$ and $y$ component, as well as the oscillations which are due to perturbing the component away from the vortex core.  This effect, combined with the non-trivial shape of the interaction potential experienced by the in-filling component from the density depletion in the majority component, lead to the in-filling component tracing out an irregular trajectory in the $x-y$ plane. It is clear that the coupled solution is stable against small perturbations of the in-filling component, and that the in-filling component remains localised within the vortex core.

\section{Conclusions}
\label{section:conclusions}
We have considered a two-component Bose Einstein Condensate which is in the immiscible regieme, where one component (the majority component) contains a vortex and the other component (the minority component resides in the vortex core. For low in-filling atom numbers, the vortex profile is not significantly different from that of a single-component vortex, while a larger number of in-filling atoms leads to a broadening and flattening of the vortex core. We have presented an ansatz for the wavefunction of each component in a uniform system based on a super-Gaussian function. Following a variational approach using this ansatz, we were able to shown that the parameters which minimise the GPE energy functional lead to wavefunctions which are in excellent agreement with the numerical solutions obtained by evolving the full coupled GPE equations for a range of atom numbers.  This approach may be extended in the future to include trapping potentials or vortex pair solutions. 

We then proceeded to consider the response of the coupled vortex solution to perturbation. We were able to ascertain that the solution is stable against perturbations of the in-filling component away from the vortex core, for a range of atom numbers and inter-species interaction strengths. 

This work was partly motivated by the prospect of using the minority, in-filling component as a passive tracer of vortex lines in atomic BECs.  Our work confirms two essential criteria for such a prospect - firstly, that the in-filling component remains localised in the vortex core (even under perturbation) and secondly, for suitably small atom numbers, has no significant affect on the vortex profile or back-action on the majority component.  Further work is needed to establish how the in-filling component behaves in more complex vortex configurations, such as three-dimensional vortex tangles.


This work was also partly motivated by whether the in-filling component can alter the vortex properties and potentially open up new physical regimes of vortex dynamics.  Indeed, the significant change to the vortex core profile for large numbers of in-filling atoms suggests a significant affect on the vortex-vortex interaction.  This interaction underpins many macroscopic vortex phenomena such as quantum turbulence, Abrikosov vortex lattices, Onsager vortex states and the Berezinskii–Kosterlitz–Thouless transition.  Studying how the in-filling component modifies the vortex-vortex interaction is an avenue for further work.

 

\acknowledgements
The authors thank Dr Thomas Bland and Dr Srivatsa Prasad for useful discussions. This work made use of the Rocket HPC facility at Newcastle University.

\appendix

\section{Integrating the Energy Functional}
\label{Appendix1}
Substituting the variational solution, Eqns.~\eqref{eqn:majority_ansatz} and \eqref{eqn:infill_ansatz}, into the energy functional, Eqn.~\eqref{eqn:energy_functional}, leads to an expression for the total energy of the solution,
\begin{equation}
    E = E_\mathrm{Kin,1a} + E_\mathrm{Kin,1b} + E_\mathrm{Kin,2}  + E_\mathrm{Int,1} + E_\mathrm{Int,12} + E_\mathrm{Int,2}, 
\end{equation}
where
\begin{subequations}
\begin{align}
    E_\mathrm{Kin,1a} &= \frac{1}{2} \int \ d^2 \rr \ \left(\frac{\partial \psi_1}{\partial r} \right)^2 , \label{eqn:E_kin_1a} \\
    E_\mathrm{Kin,1b} &= \frac{1}{2} q^2 \int \ d^2 \rr \ \frac{1}{r^2} \left( \psi_1 \right)^2 , \label{eqn:E_kin_1b}\\
    E_\mathrm{Kin,2} &= \frac{1}{2} m' \int \ d^2 \rr \  \left(\frac{\partial \psi_2}{\partial r}\right)^2, \label{eqn:E_kin_2}\\
    E_\mathrm{Int,1} &= \frac{1}{2} \int \ d^2 \rr \ |\psi_1|^4, \label{eqn:E_int_1}\\ 
    E_\mathrm{Int,12}  &= g_{12}^\prime \int \ d^2 \rr \ |\psi_1|^2 |\psi_2|^2, \label{eqn:E_int_12}\\
    E_\mathrm{Int,2} &= \frac{1}{2} g_{22}^\prime \int \ d^2 \rr \ |\psi_2|^4. \label{eqn:E_int_2}
\end{align}
\end{subequations}
We give brief details on computing these in the following subsections. 
\subsection{Kinetic Terms}
In order to compute the kinetic terms, Eqns.~\eqref{eqn:E_kin_1a}--\eqref{eqn:E_kin_2}, we must find the gradients of $\psi_1$ and $\psi_2$ which are given by 
\begin{eqnarray}
    \nabla \psi_1 &=& \frac{A \alpha}{r} \left(\frac{r}{\lambda}\right)^{2\alpha} \exp\left[-\left(\frac{r}{\lambda}\right)^{2\alpha}\right] \left\{ 1 - \exp\left[-\left(\frac{r}{\lambda}\right)^{2\alpha}\right]\right\}^{-1/2} e^{iq\varphi} \  \mathbf{\hat{r}} \nonumber \\
    &+& \frac{A i q}{r^2} \left\{ 1 - \exp\left[-\left(\frac{r}{\lambda}\right)^{2\alpha}\right]\right\}^{1/2}e^{iq\varphi} \ \mathbf{\hat{\varphi}}
\end{eqnarray}
and 
\begin{equation}
    \nabla \psi_2 = \frac{B\alpha}{r} \left(\frac{r}{\lambda}\right)^{2\alpha} \exp\left[-\frac{1}{2}\left(\frac{r}{\lambda}\right)^{2\alpha}\right] \ \mathbf{\hat{r}}.
\end{equation}
The first integral, Eqn.~\eqref{eqn:E_kin_1a}, is then given by
\begin{eqnarray}
    E_\mathrm{Kin,1a} &=& \pi A^2 \alpha^2 \int_0^R   \dfrac{ \frac{1}{r} \left(\frac{r}{\lambda}\right)^{4\alpha}  \exp\left[-2\left(\frac{r}{\lambda}\right)^{2\alpha}\right]}{ 1 - \exp\left[-\left(\frac{r}{\lambda}\right)^{2\alpha}\right]} \ dr, \nonumber \\
    &=& \frac{\pi A^2 \alpha^2}{2} \int_{u(0)}^{u(R)} \frac{u e^{-2u}}{1-e^{-u}} \ du,
\end{eqnarray}
where the second line is obtained by making the substitution $u(r)=\left(r/\lambda\right)^{2\alpha}$. The solution is 
\begin{equation}
    E_\mathrm{Kin,1a} = \frac{\pi A^2 \alpha^2}{2} \left\{ \left[1 + u(R) \right] \exp\left[-u(R)\right] + \mathrm{Sp}\left[e^{-u(R)} \right] - 1 \right\},
\end{equation}
where we have introduced the Spence function \cite{Abramowitz1965handbook}, which is defined as
\begin{equation}
    \mathrm{Sp}(x) = - \int_1^x \frac{\log t}{t-1} \ dt.
    \label{eqn:Spence_integral}
\end{equation}
The second term is the kinetic energy of the azimuthally circulating fluid motion \cite{Pethick_and_Smith}, which here is given by
\begin{equation*}
    E_\mathrm{Kin,1b} = A^2 \pi q^2 \int_0^R \frac{1}{r} \left\{ 1 - \exp\left[-\left(\frac{r}{\lambda}\right)^{2\alpha}\right]\right\} \ dr.
\end{equation*}
The Maclaurin series expansion of the integrand confirms that this integral does not diverge at the origin. Then it can be shown that 
\begin{eqnarray}
\int_\varepsilon^R \frac{1}{r} \left\{ 1 - \exp\left[-\left(\frac{r}{\lambda}\right)^{2\alpha}\right]\right\} \ dr = \left\{ \log\left(\frac{r}{\lambda}\right) - \frac{1}{2\alpha} \mathrm{Ei}\left[-\left(\frac{r}{\lambda}\right)^{2\alpha}\right]\right\} \bigg{|}^R_\varepsilon, \nonumber \\
\label{eqn:formal_azimuthal_integral}
\end{eqnarray}
where we have defined the exponential integral $\mathrm{Ei}(x)$ as 
\begin{equation}
    \mathrm{Ei}(x) = \int_x^\infty \frac{e^{-t}}{t} dt. 
    \label{eqn:Exponential_integral}
\end{equation}
It is possible to consider the analytic continuation of $\mathrm{Ei}(x)$ along the negative real axis \cite{Abramowitz1965handbook}, given by $\mathrm{E}_1(x)=-\mathrm{Ei}(-x)$ which leads to the series expansion
\begin{equation}
    \mathrm{Ei}(-x) = \gamma_\mathrm{EM} + \log x + \sum_{n=1}^\infty \frac{(-1)^n}{n} \frac{x^n}{n!},
\end{equation}
where $\gamma_\mathrm{EM}$ is the Euler-Mascheroni constant, $\gamma_\mathrm{EM}\approx 0.5772$. Then the lower limit of the integral in Eqn.~\eqref{eqn:formal_azimuthal_integral} is
\begin{eqnarray*}
- \log \left(\frac{\varepsilon}{\lambda}\right) + \frac{1}{2\alpha}\left\{ \gamma_\mathrm{EM} + \log\left[ \left(\frac{\varepsilon}{\lambda}\right)^{2\alpha}\right] + \sum_{n=1}^\infty \frac{(-1)^n}{n n!} \left[\left(\frac{\varepsilon}{\lambda}\right)^{2\alpha}\right]^n \right\},
\end{eqnarray*}
which tends to $\gamma_\mathrm{EM}/2\alpha$ as $\varepsilon\to 0$, since the final term is a sum of positive powers of $\varepsilon$. Hence, the resulting form of the kinetic energy of the azimuthal motion is  
\begin{equation}
    E_\mathrm{Kin,1b} = \frac{\pi q^2 A^2}{2\alpha}\left\{ \log\left[\left(\frac{R}{\lambda}\right)^{2\alpha}\right] - \mathrm{Ei}\left[-\left(\frac{R}{\lambda}\right)^{2\alpha}\right] + \gamma_\mathrm{EM} \right\}.
    \label{eqn:E_kin_1b_result}
\end{equation}

The final kinetic term is due to the infill component, and is given by
\begin{equation}
    E_\mathrm{Kin,2} = \pi m B^2 \alpha^2 \int_0^R \frac{1}{r} \left(\frac{r}{\lambda}\right)^{4\alpha} \exp\left[-\left(\frac{r}{\lambda}\right)^{2\alpha}\right] \ dr .
\end{equation}
This is readily computed on making the substitution $u(r)=(r/\lambda)^{2\alpha}$, resulting in
\begin{equation}
    E_\mathrm{Kin,2} = \frac{\pi m B^2 \alpha^2}{2} \left\{ 1 - \left[ 1 + \left(\frac{R}{\lambda}\right)^{2\alpha}\right]\exp\left[-\left(\frac{R}{\lambda}\right)^{2\alpha}\right]\right\}. 
        \label{eqn:E_kin_2_result}
\end{equation}


\subsection{Interaction Terms}
The final terms to compute are due to the interaction terms of the energy functional, Eqns.~\eqref{eqn:E_int_1}--\eqref{eqn:E_int_2}. It's possible, however, to save some work in noticing that these integrals contain terms of three forms:
\begin{subequations}
\begin{align}
   \int_0^R \, r \, dr = \frac{1}{2} R^2, \\
\int_0^R \, r \exp\left[-\left(\frac{r}{\lambda}\right)^{2\alpha}\right] \, dr = \frac{\lambda^2}{2\alpha}\Gamma\left[\frac{1}{\alpha},\left(\frac{R}{\lambda}\right)^{2\alpha}\right], \\
\int_0^R \, r \exp\left[-2\left(\frac{r}{\lambda}\right)^{2\alpha}\right] \, dr = \frac{\lambda^2}{2^{1 + 1/\alpha} \alpha}\Gamma\left[\frac{1}{\alpha},2\left(\frac{R}{\lambda}\right)^{2\alpha}\right], 
\end{align}
\end{subequations}
where we have introduced the incomplete Gamma function 
\begin{equation}
    \Gamma\left(s,x\right) = \int_0^x t^{s-1} e^{-t} \ dt.
    \label{eqn:incomplete_gamma_function}
\end{equation}
which is often referred to the \textit{lower} incomplete Gamma function, see for example \cite{Abramowitz1965handbook}. Then the terms in the energy functional are as follows: firstly,
\begin{eqnarray}
E_\mathrm{Int,1} &=& \pi A^4 \int_0^R \ r \left\{ 1 - \exp\left[-\left(\frac{r}{\lambda}\right)^{2\alpha}\right]\right\}^2 \ dr \nonumber \\ 
&=& \frac{\pi A^4}{2} \left\{ R^2 - \frac{2 \lambda^2}{\alpha} \Gamma\left[\frac{1}{\alpha},\left(\frac{R}{\lambda}\right)^{2\alpha}\right] + \frac{\lambda^2}{2^{1/\alpha} \alpha} \Gamma\left[\frac{1}{\alpha},2\left(\frac{R}{\lambda}\right)^{2\alpha}\right] \right\}, \nonumber \\
\label{eqn:E_int_1_result}
\end{eqnarray}
secondly,
\begin{eqnarray}
 E_\mathrm{Int,12} &=&  2 \pi g_{12} A^2 B^2 \int_0^R \, r \left\{1- \exp\left[-\left(\frac{r}{\lambda}\right)^{2\alpha}\right] \right\} \exp\left[-\left(\frac{r}{\lambda}\right)^{2\alpha}\right] \, dr \nonumber \\
&=& \pi g_{12} A^2 B^2 \left\{\frac{\lambda^2}{\alpha}\Gamma\left[\frac{1}{\alpha},\left(\frac{R}{\lambda}\right)^{2\alpha}\right] - \frac{\lambda^2}{2^{1/\alpha} \alpha}\Gamma\left[\frac{1}{\alpha},2\left(\frac{R}{\lambda}\right)^{2\alpha}\right]\right\}, \nonumber \\
\label{eqn:E_int_12_result}
\end{eqnarray}
and finally, 
\begin{eqnarray}
& \ & E_\mathrm{Int,2} = \nonumber \\
& \ & \pi g_{22} B^4 \int_0^R \exp\left[-2\left(\frac{r}{\lambda}\right)^{2\alpha}\right] = \frac{\pi g_{22} B^4 \lambda^2}{2^{1 + 1/\alpha} \alpha}\Gamma\left[\frac{1}{\alpha},2\left(\frac{R}{\lambda}\right)^{2\alpha}\right]. \nonumber \\
\label{eqn:E_int_2_result}
\end{eqnarray}
The resulting energy functional is 
\begin{eqnarray}
& \ & E = \nonumber \\
& \ & \frac{\pi A^2 \alpha^2}{2} \left( \left[1 + \left(\frac{R}{\lambda}\right)^{2\alpha} \right] \exp\left[-\left(\frac{R}{\lambda}\right)^{2\alpha}\right] + \mathrm{Sp}\left\{\exp\left[-\left(\frac{R}{\lambda}\right)^{2\alpha}\right] \right\}\right) \nonumber \\
&+& \frac{\pi q^2 A^2}{2\alpha}\left\{ \log\left[\left(\frac{R}{\lambda}\right)^{2\alpha}\right] - \mathrm{Ei}\left[-\left(\frac{R}{\lambda}\right)^{2\alpha}\right]\right\} \nonumber \\
&+& \frac{\pi m B^2 \alpha^2}{2} \left\{ 1 - \left[ 1 + \left(\frac{R}{\lambda}\right)^{2\alpha}\right]\exp\left[-\left(\frac{R}{\lambda}\right)^{2\alpha}\right]\right\} \nonumber \\
&+& \frac{\pi A^4}{2} \left\{ R^2 - \frac{2 \lambda^2}{\alpha} \Gamma\left[\frac{1}{\alpha},\left(\frac{R}{\lambda}\right)^{2\alpha}\right] + \frac{\lambda^2}{2^{1/\alpha} \alpha} \Gamma\left[\frac{1}{\alpha},2\left(\frac{R}{\lambda}\right)^{2\alpha}\right] \right\} \nonumber \\
&+& \pi g_{12} A^2 B^2 \left\{\frac{\lambda^2}{\alpha}\Gamma\left[\frac{1}{\alpha},\left(\frac{R}{\lambda}\right)^{2\alpha}\right] - \frac{\lambda^2}{2^{1/\alpha} \alpha}\Gamma\left[\frac{1}{\alpha},2\left(\frac{R}{\lambda}\right)^{2\alpha}\right]\right\} \nonumber \\
&+&\frac{\pi g_{22} B^4 \lambda^2}{2^{1 + 1/\alpha} \alpha}\Gamma\left[\frac{1}{\alpha},2\left(\frac{R}{\lambda}\right)^{2\alpha}\right]. 
\end{eqnarray}


\bibliography{two-component_variational_paper}
\end{document}